\documentclass[aps,prl,reprint,groupedaddress]{revtex4-1}

\usepackage{amsmath,amssymb,graphicx}
\usepackage{array}
\usepackage[caption=false]{subfig}
\usepackage{bm}
\usepackage{amsfonts}
\usepackage{epsfig,epstopdf}
\usepackage{subfig}
\usepackage{url}

\def\Tr{{\rm Tr}}

\def\la{\left\langle}
\def\ra{\right\rangle}

\def\br{{\bf r}}
\def\bq{{\bf q}}
\def\Prob{{\rm Prob}}

\def\cR{{\cal R}}

\def\bE{\mathbb{E}}

\def\bJ{{\bf J}}

\def\bu{{\bf u}}

\def\hrho{\hat \rho}

\def\bH{{\bf H}}

\def\be{\begin{equation}}
\def\ee{\end{equation}}
\def\ba{\begin{align}}
\def\nn{\nonumber\\}

\def\defeq{\buildrel \rm def \over =}

\def\bs{{\bf s}}
\def\b0{{\bf 0}}
\def\cA{{\cal A}}
\def\la{\left\langle}
\def\ra{\right\rangle}

\def\br{{\bf r}}
\def\bs{{\bf s}}

\def\bu{{\bf u}}
\def\bE{\mathbb{E}}

\def\bJ{{\bf J}}

\def\bH{{\bf H}}

\def\bl{{\boldsymbol \ell}}
\def\blperp{{\bm{l}}_\perp}

\def\cO{{\cal O}}

\def\Tr{{\rm Tr\ }}

\def\be{\begin{equation}}
\def\ee{\end{equation}}
\def\ba{\begin{align}}
\def\nn{\nonumber\\}
\def\la{\langle}
\def\ra{\rangle}

\def\cX{{\cal X}}
\def\CC{{\rm CC}}
\def\CS{{\rm CS}}
\def\SC{{\rm SC}}
\def\SS{{\rm SS}}
\def\defeq{\buildrel \rm def \over =}

\begin{document}

\title{Quantum limited superresolution of an incoherent source pair in three dimensions}

\author{Zhixian Yu}
\author{Sudhakar Prasad}
\altaffiliation{Also at School of Physics and Astronomy, University of Minnesota, Minneapolis, MN 55455}

\email{sprasad@unm.edu}
\affiliation{Department of Physics
and Astronomy, University of New Mexico, Albuquerque, New Mexico 87131}
\date{\today}


\begin{abstract}

The error in estimating the separation of a pair of incoherent sources from radiation emitted by them and subsequently captured by an imager is fundamentally bounded below by the inverse of the corresponding quantum Fisher information (QFI) matrix. We calculate the QFI for estimating the full three-dimensional (3D) pair separation vector,  extending previous work on pair separation in one and two dimensions.  We also show that the pair-separation QFI is, in fact, identical to source localization QFI, which underscores the fundamental importance of photon-state localization in determining the ultimate estimation-theoretic bound for both problems. We also propose general coherent-projection bases that can attain the QFI in two special cases. We present simulations of an approximate experimental realization of such quantum limited pair superresolution using the Zernike basis, confirming the achievability of the QFI bounds. 
\end{abstract}

\vspace{-1cm}


\maketitle

Rayleigh's pair-resolution criterion\cite{Rayleigh} is routinely superseded by modern imaging systems. An approach that entirely circumvents it employs PSF fitting and localization of single fluorescent molecules by selective excitation in which two closeby molecules are rarely, if ever, excited simultaneously \cite{Moerner89, Hell94, Betzig95} in each frame, thus allowing a frame-by-frame construction of a composite, superresolved image of a collection of densely packed molecules. Another, more direct approach uses computational image processing with {\it a priori} constraints under sufficiently high pixel brightness \cite{RH68,BdM96,Lucy92,SM04, RWO06, Prasad14}.

The covariance matrix, $V_\theta[\cO,\check\theta]$, for the unbiased estimator, $\check\theta$, of a set of quantities, $\theta\defeq \{\theta_p\mid \ p=1,\ldots,P\}$, parameterizing the density operator, $\hat\rho_\theta$, of a system is bounded below by the inverse of the quantum Fisher information (QFI) matrix \cite{Helstrom76,Braunstein94,Paris09,Szczykulska16,Safranek18}, namely the quantum Cram\'er-Rao bound (QCRB),
\be
\label{info_ineq}
V_\theta[\cO,\check\theta] \geq \bJ^{-1}_\theta[\cO]\geq \bH^{-1}_\theta,
\ee
in which $\cO=\{\hat O(x)\mid x\in \cX\}$ defines a positive-operator valued measure (POVM) of non-negative operators defined on a data set $\cX$ and which sum to the identity operator, $\hat I$. 
The classical FI matrix, $\bJ_\theta[\cO]$, is defined \cite{VT68,Kay93} in terms of the probability distribution (PD) of the POVM, $P_O(x;\theta)=\Tr [\hat\rho_\theta \hat O(x)]$, as 
\be
\label{CFI}
\bJ_\theta [\cO]= \bE_O\left(\nabla_\theta \ln P_O(x;\theta)\nabla^T_\theta\ln P_O(x;\theta)\right),
\ee
in which $\nabla_\theta \ln P$ is a column vector representing the gradient taken relative to $\theta$, the superscript $T$ denotes matrix transpose, and $\bE_O$ the statistical expectation of its argument over the PD. The inverse of the classical FI is the classical Cram\'er-Rao lower bound (CRB).

Tsang {\it et al.} \cite{Tsang16} proved that pair separation can achieve QCRB in one dimension with classical wavefront projections. This has been generalized to a thermal source pair of the same average but otherwise indefinite strength  \cite{Nair16}, to a source pair in an arbitrary quantum state \cite{Lupo16}, to homodyne and heterodyne detection\cite{Yang17}, and to two dimensions \cite{Ang17}, and experimentally verified by a number of groups \cite{Paur16,Tang16,Yang16,Tham17}. For an imager with a one dimensional (1D) Gaussian point-spread function (PSF), it is the Hermite Gaussian (HG) basis \cite{Tsang16} that perfectly achieves QCRB, which turns out to be independent of the pair separation. By contrast, the conventional image-based approach entails a quadratic dependence of FI on the separation. This critical difference implies dramatically different inverse-square vs. inverse-quartic power-law scalings of the minimum photon number needed to resolve the pair as a function of their separation using these two approaches. 

Here we treat the problem of estimating the full 3D separation vector for a pair of incoherent, equally bright point sources, when the pair centroid is known and an imager with a circular aperture is used \cite{PrasadYu17}. We first calculate the $3\times 3$ QFI matrix with respect to (w.r.t.) the three components of the pair separation vector, and show it to be diagonal and independent of the latter. We also show that QFI is in fact the same as that for localizing a single point emitter in 3D \cite{Backlund18}. We then discuss projective-measurement protocols that can achieve QCRB in two special cases of vanishing axial and lateral separations. We finally present simulations of an experimental proposal to achieve quantum-limited 3D pair separation. 

A photon emitted by an incoherent pair of equally bright point sources  is described by the density operator,
\be
\label{e0}
\hat\rho = {1\over 2}\left(|K_+\ra\la K_+| + |K_-\ra\la K_-|\right),
\ee
in which $|K_\pm\ra$ are pure one-photon states corresponding to its emission individually by the two sources taken to be at the 3D locations, $\pm (\br_\perp,r_z)$, with respect to their centroid. The corresponding normalized transverse and axial semi-separations, $\blperp,l_z$, are defined as
\be
\label{norm_separation}
\bl_\perp = \br_\perp/\sigma_0, \ \ l_z=r_z/\zeta_0,
\ee
where $\sigma_0=\lambda z_O/R$ and $\zeta_0=\lambda z_O^2/R^2$ denote the characteristic transverse and axial resolution scales  for an aperture of radius $R$, optical wavelength $\lambda$, and distance $z_O$ of the pair centroid from the aperture\cite{Goodman17}.  

The coordinate representations, $\la\bs|K_\pm\ra$, of these states are the corresponding image-plane amplitude PSFs. Their momentum-space representations are the corresponding wavefunctions in the exit pupil of the imager \cite{Goodman17}, 
\be
\label{e1}
\la\bu|K_\pm\ra
=\exp(\pm i\phi_0)\, P(\bu)\exp[\mp i (2\pi \blperp\cdot\bu+\pi l_zu^2)],
\ee
in which $P(\bf u)$ denoted a general aperture function. For a clear aperture, $P(\bu)$  is simply $1/\sqrt{\pi}$ times its indicator function, corresponding to the Airy PSF, while in its Gaussian form, it yields the Gaussian PSF. Most generally, $P(\bu)$ need only obey the normalization condition,
\be
\label{norm1}
\int d^2u\, |P(\bu)|^2 =1,
\ee
that follows from requiring $\la K_\pm |K_\pm\ra=1$.

The two non-zero eigenvalues, $e_\pm$, and the associated orthonormal eigenstates, $|e_\pm\ra$, of $\hrho$ given by Eq.~(\ref{e0}) are 
\be
\label{eigen}
e_\pm = (1\pm \Delta)/ 2;\ \ |e_\pm\ra =[2(1\pm \Delta)]^{-1/2}\left(|K_+\ra\pm |K_-\ra\right),
\ee
where $\Delta$ is the inner product, $\Delta=\la K_-|K_+\ra,$
which we render real and positive by a proper choice of the phase constant, $\phi_0$.

The QFI matrix has elements, ${\rm Re}H_{\mu\nu}$, where Re denotes the real part and $H_{\mu\nu}\defeq \Tr(\hrho \hat L_\mu\hat L_\nu)$ can be expressed \cite{supp} in the eigenbasis of $\hrho$ as 
\be
\label{Hmn}
H_{\mu\nu}=\sum_{i\in\cR} \sum_j {4e_i\over (e_i+e_j)^2} 
\langle e_i| \partial_\mu \hat\rho|e_j\rangle\langle e_j| \partial_\nu \hat\rho|e_i\rangle, 
\ee
in which $\hat L_\mu$ is the symmetric logarithmic derivative (SLD) of $\hrho$ w.r.t. parameter $l_\mu$, for brevity we denote $\partial \hat\rho/\partial l_\mu$ as $\partial_\mu\hat\rho$, and $\cR$ denotes the set of values of an index for the eigenstates that span the range space of $\hrho$.
 
By decomposing the $j$ sum into a sum over the range space of $\hat\rho$ and another over its null space, $j\notin \cR$ for which $e_j=0$, we may evaluate the latter sum via the completeness relation, 
$$\sum_{j\notin \cR}|e_j\rangle\langle e_j|=\hat I-\sum_{j \in \cR}|e_j\rangle\langle e_j|.$$
We may thus express $H_{\mu\nu}$ in Eq.~(\ref{Hmn}) as 
\begin{align}
\label{Hmn1}
H_{\mu\nu}&=\sum_{i\in \cR}{4\over e_i}\langle e_i|\partial_\mu \hat\rho\partial_\nu \hat\rho|e_i\rangle\nn
&+\sum_{i\in \cR}\sum_{j\in \cR}\left[{4e_i\over {(e_i+e_j)}^2}-{4\over e_i}\right]\langle e_i|\partial_\mu \hat\rho|e_j\rangle\langle e_j|\partial_\nu \hat\rho|e_i\rangle.
\end{align}
 
For the present problem for which $\cR =\{+,-\}$, we may simplify the derivatives in Eq.~(\ref{Hmn1}) by means of the eigenvector identity, $\partial_\mu [(\hat\rho-e_i\hat I)|e_i\rangle]=0,$ 
and thus express  $H_{\mu\nu}$ as \cite{supp}
\begin{align}
\label{Hmn2}
H_{\mu\nu}&=\sum_{i=\pm}{1\over e_i}\partial_\mu e_i\partial_\nu e_i+4\sum_{i=\pm}{1\over e_i}(\partial_\mu \langle 
e_i|)(\hat\rho-e_i\hat I)^2\partial_\nu|e_i\rangle\nn
&+4\Delta^2\sum_{i\neq j}\left({1\over e_i}-e_i\right)\langle e_i|\partial_\mu |e_j\rangle\langle e_j|\partial_\nu |e_i\rangle,
\end{align}
in which we used the identities, $e_++e_-=1$ and $e_+-e_-=\Delta$.
The first sum in expression (\ref{Hmn2}) may be regarded as the classical part of QFI, the real part of the second sum the contribution of quantum fluctuations of the photon state to QFI, and the real part of the final sum an additional contribution  from the pair cross-coherence, $\Delta\neq 0$. 

By evaluating the various state derivatives in expression (\ref{Hmn2}), we may reduce it further \cite{supp} to the form,
\be
\label{Hmn4}
H_{\mu\nu}
=4\left[(\partial_\mu  \langle K_+|)\partial_\nu|K_+\rangle+\langle K_+|\partial_\mu |K_+\rangle \langle K_+|\partial_\nu|K_+\rangle\right].
\ee
By using expression (\ref{e1}) for $\la\bu|K_+\ra$, we may evaluate Eq.~(\ref{Hmn4}) in terms of the gradient of the phase function,
\be
\label{Phase}
\Psi(\bu;\bl) = 2\pi \blperp\cdot \bu +\pi l_z u^2,
\ee
independently of $\phi_0$ as
\be
\label{Hmn5}
H_{\mu\nu}
=4\left[\langle \partial_\mu\Psi\partial_\nu\Psi\rangle-\langle \partial_\mu\Psi\rangle\langle\partial_\nu\Psi\rangle\right],
\ee 
where angular brackets now denote averages over the modulus squared aperture function, $|P(\bu)|^2$. 

Form (\ref{Hmn5}) of QFI underscores the fundamental role of the correlations of the wavefront gradient in the aperture in controlling the error of estimation of the pair separation. For a clear circular aperture, to which we restrict attention in the rest of the paper and for which $|P(\bu)|^2$ is $1/\pi$ times its indicator function, simple integrations yield the following averages:
\be
\label{phase_corr}
\langle u_i\rangle=0; \ \langle u_i u_j\rangle={\delta_{ij} \over 4}; \ \langle u^2\rangle={1\over 2};\ \langle u^4\rangle ={1\over 3};\ \ i,j=x,y,
\ee
and thus the following purely diagonal form of the per-photon 3D QFI matrix:
\be
\label{H}
\bH(l_x, l_y, l_z) = \left(
\begin{array}{lll}
\displaystyle{4\pi^2} & 0& 0\\
0& \displaystyle{4\pi^2}& 0 \\
0&0&\displaystyle{\pi^2\over 3}
\end{array}
\right).
\ee
The reality and diagonal character of $H_{\mu\nu}$ provide necessary and sufficient achievability conditions for the simultaneous estimation of the three separation coordinates in the asymptotic limit \cite{Ragy16}.  

We next show that QFI  for localizing a single source, say the one located at $+(\blperp,l_z)$, is identical to that we have just obtained for 3D pair separation. For this problem, only the middle term in expression (\ref{Hmn2}) contributes, since $\hrho=|K_+\ra\la K_+|$ has a single fixed non-zero eigenvalue, $e_+=1$, with eigenstate $|e_+\ra=|K_+\ra$, and $(\hrho-\hat I)^2=\hat I-|K_+\ra\la K_+|$. In view of these relations and normalization, $\la K_+|K_+\ra=1$, which requires that $(\partial_\mu\la K_+|)|K_+\ra=-\la K_+|\partial_\mu|K_+\ra$, the resulting QFI becomes identical to Eq.~(\ref{Hmn4}) for QFI for source-pair separation. The 3D source-localization QFI has been calculated directly from the definition of SLD of the density opermator for a pure state in Ref.~\cite{Backlund18}, but unlike that approach ours can be more efficiently extended, numerically if necessary, to calculate QFI for joint source localization and separation of two or more sources emitting single photons in a general state \cite{PrasadYu18}. The equality of the QFI matrices for source localization and pair separation shows that the general problem is one of estimating the photon state, independent of the nature of its emitter.

QCRB is achievable via orthonormal wavefront projections in two special cases. For sources in the same transverse plane, for which $l_z=0$, consider an orthonormal basis, $\cA=\{A_{mn}(\bu)|m,n\in \mathbb{Z}\}$, of states in the aperture plane obeying the condition, $|\la K_+|A_{mn}\ra|=|\la K_-|A_{mn}\ra|, \ \forall m,n$. Since $\la\bu|K_+\ra=\la\bu|K_-\ra^*$, this condition is met by {\it any} real basis. The probability $P^{(A)}_{mn}$, which is equal to $\la A_{mn}|\hrho|A_{mn}\ra$, may then be written as $P^{(A)}_{mn} = |\la K_+|A_{mn}\ra|^2,$ from which follow the FI matrix elements,
\ba
\label{FI_A}
J_{\mu\nu}[\cA]=&\sum_{m,n} {\partial_\mu P^{(A)}_{mn}\, \partial_\nu P^{(A)}_{mn}\over P^{(A)}_{mn}}\nn
                      =&4\sum_{m,n} \partial_\mu|\la A_{mn}|K_+\ra|\, \partial_\nu |\la A_{mn}|K_+\ra|.
\end{align}
If we assume further that the phases of $\la K_+|A_{mn}\ra$ are independent of $\blperp$, then Eq.~(\ref{FI_A}) simplifies to                 
\ba
\label{FI_A2}
J_{\mu\nu}[\cA]=&4\sum_{m,n}(\partial_\mu\la K_+|) |A_{mn}\ra\la A_{mn}|\partial_\nu|K_+\ra\nn
                      =&4(\partial_\mu\la K_+|)\partial_\nu|K_+\ra,
\end{align}
with the second equality following from the completeness relation, $\sum_{m,n} |A_{mn}\ra\la A_{mn}|=\hat I$.
For $\mu,\nu=x,y$, $J_{\mu\nu}[\cA]$ matches QFI in expression (\ref{Hmn4}) since for the choice, $\phi_0=0$, we make to render the phases of $\la K_+|A_{mn}\ra$ independent of $\blperp$, $\la K_+|\partial_\mu|K_+\ra$, vanishes identically for any inversion symmetric aperture.

The orthonormal sine-cosine Fourier basis states in polar coordinates, $(u,\phi)$,
\be
\label{basis1}
\begin{array}{ll}
\CC_{mn}(\bu)=\sqrt{c_m c_n\over \pi}\cos (2\pi m u^2)\cos n\phi,  & m,n=0,1,\ldots; \\
\CS_{mn}(\bu)=\sqrt{c_m c_n\over \pi}\cos (2\pi m u^2)\sin n\phi,  & m=0,1,\ldots,\\
&\ n=1,2,\ldots; \\
\SC_{mn}(\bu)=\sqrt{c_m c_n\over \pi}\sin (2\pi m u^2)\cos n\phi,  & m=1,2,\ldots,\\
&\ n=0,1,\ldots \\
\SS_{mn}(\bu)=\!\sqrt{c_m c_n\over \pi}\sin (2\pi m u^2)\sin n\phi,  & m,n=1,2,\ldots;
\end{array}
\ee 
with $c_n=2-\delta_{n0}$, constitute one such basis that achieves QFI for the case of pure transverse pair separation as their overlap integrals with the photon wavefront of each source can be readily shown \cite{supp} to have phases that are independent of that separation. 

For the source pair being on the optical axis, {\it i.e.}, $l_\perp=0$, only the $n=0$ subset of the sine-cosine basis, as we need no angular localization, achieves QCRB w.r.t. $l_z$, as we show next. The relevant probability amplitudes are 
 \ba
\label{prob_amp}
\la &A_{m0}|K_+\ra={1\over\sqrt{\pi}}\int_0^1 du\, u \exp(-i\pi l_z u^2)\, A_{m0}(u);\nn
                            =&{1\over 2\sqrt{\pi}}\exp\left(-i\pi {l_z\over 2}\right)\int_{-1/2}^{1/2}\!\!dv \cos(\pi l_z v) A_{m0}(\sqrt{v+1/2}),
\end{align}
with $A=\CC,\SC$. We used the variable transformation, $v=u^2-1/2$, followed by a symmetrization of the resulting integrand to reach the second equality in Eq.~(\ref{prob_amp}) that involves a purely real integral. In view of the form (\ref{prob_amp}), we have $|\la A_{m0}|K_+\ra| = \exp(i\pi l_z/2)\la A_{m0}|K_+\ra$, which allows us, analogously to Eq.~(\ref{FI_A}) with $\mu=\nu=z$, to express FI w.r.t. $l_z$ as
\ba
\label{FI_zz}
J_{zz}[\cA]=& 4\sum_m \big|\partial_z |\la A_{m0}|K_+\ra|\big|^2\nn
                 =&4\sum_m \big[\partial_z(\la K_+|)|A_{m0}\ra -i(\pi/2)\la K_+|A_{m0}\ra\big]\nn
                  &\times \big[\la A_{m0}|\partial_z|K_+\ra +i(\pi/2)\la A_{m0}|K_+\ra\big]\nn
                  =&4\big[\partial_z(\la K_+|)|\partial_z|K_+\ra -i(\pi/2)\la K_+|\partial_z|K_+\ra\nn
                  &+i(\pi/2)(\partial_z|K_+)|K_+\ra +(\pi/2)^2\big]\nn
                  =&4\big[\partial_z(\la K_+|)|\partial_z|K_+\ra -\pi^2/4\big]\nn
                  =&4\big[\partial_z(\la K_+|)|\partial_z|K_+\ra +\la K_+|\partial_z|K_+\ra^2\big]
\end{align}
in which we used the completeness of the $|A_{m0}\ra$ states over the aperture for $\phi$-invariant wavefunctions like $\la \bu|K_+\ra$ characteristic of an axially separated source pair and relations, $\la K_+|\partial_z|K_+\ra=(\partial_z\la K_+|)|K_+\ra^*=-i\pi\la u^2\ra=-i\pi/2$, to derive the various expressions. We see from expression (\ref{Hmn4}) that the $\{A_{m0}|A=\CC,\SC,\ m=0,1,\ldots\}$ basis achieves QFI w.r.t. $l_z$ for an axially separated source pair.  More generally, {\it any real} basis of orthonormal projections, $\{|B_m\ra \}$, for which the equality, $|\la B_m|K_+\ra|=|\la B_m|K_-\ra|$, certainly holds, will achieve QFI.

Projections that are well matched to the linear tilt and quadratic defocus parts of the aperture phase function, $\Psi(\bu)$, given by Eq.~(\ref{Phase}), can achieve full 3D QFI in the limit of small separations, $l_\perp, l_z << 1$.  One need merely use a few such projections, as noted in Ref.~\cite{Tsang16}), to attain quantum-limited estimation variance in this limit.  In the 3D case, we consider aperture-plane wavefront projections into low-order orthonormal Zernike basis functions \cite{Noll76}, $\{Z_n,\ n=1,2,\ldots, N\}$, with $N\sim 4-7$. Here we only discuss projections into the first four Zernikes,
\ba
\label{Zernikes}
Z_1=&{1\over \sqrt{\pi}},\  Z_2={2 \over\sqrt{\pi}}u\cos\phi,\ Z_3= {2 \over\sqrt{\pi}}u\sin\phi,\nn
Z_4=&\sqrt{3\over\pi} (2u^2-1).
\end{align}
The second and third of these correlate perfectly, respectively, with the tilt phases corresponding to the $x$ and $y$ components of the transverse separation vector, $\blperp$, and may thus be regarded as matched filters \cite{Turin60} for the latter. 
By contrast, the first and last are only partially matched to the quadratic pupil phase corresponding to the axial separation, $l_z$, with their probabilities remaining finite when $l_z\to 0$. The imperfect match of the latter with a single projection mode causes striking differences, as we shall see, in the estimation error bounds that are achievable in the limit of vanishing separation.  

We now prove these assertions by evaluating  \cite{supp} the mode-projection probabilities, $P_n=\la Z_n|\hat\rho|Z_n\ra$, 
for $l_\perp,l_z<<1$,  
\be
\label{prob10}
P_n=\left\{
\begin{array}{ll}
1-\pi^2(l_\perp^2+l_z^2/12)+O(l_\perp^4,l_z^4) & n=1\\
\pi^2 l_x^2[1+O(l_\perp^2,l_z^2)] & n=2\\
\pi^2 l_y^2 [1+O(l_\perp^2,l_z^2)]& n=3\\
\pi^2 l_z^2/12 +O(l_\perp^4, l_z^2l_\perp^2,l_z^4)& n=4
\end{array}
\right.
\ee
Since $(\partial_x P_2)^2/P_2=(\partial_y P_3)^2/P_3=4\pi^2[1+O(l_z^2)]$, we see that each reaches QFI in the limit $l_z\to 0$. By contrast, the $Z_4$ projection contributes to FI w.r.t. $l_z$ the term, $(\partial_z P_4)^2/P_4$, which is of form $(\pi^2/3)\{l_z^2/[l_z^2(1+O(l_\perp^2))+O(l_\perp^4)]\}$ and vanishes in the limit $l_z\to 0$ if $l_\perp\neq 0$. The same form implies, however, that for $l_\perp <<1$, FI as a function of $l_z$  rises to a value comparable to the QFI, $\pi^2/3$, over an interval of order $l_\perp^2$. All other contributions to the various matrix elements of FI are negligibly small in the limit of vanishing $\bl$, so the inverse of the diagonal elements of FI determine the corresponding CRBs to the most significant order in $\bl$. 

One can perform wavefront projections by digital holography \cite{Paur16}. Specifically, consider encoding the sum, $\sum_{n=1}^N Z_n(\bu)\cos (\bq_n\!\cdot\!\bu)$, 
as the distribution of the amplitude transmittance of a plate, with negative values in the sum realized by a $\pi$ phase retardation. 
Let the imaging wavefront, which is an incoherent superposition of the photon wavefunctions $\la \bu|K_\pm\ra$ and carries $M$ photons, be incident on such a plate that is placed in the aperture (or a conjugate plane thereof), and then optically focused on a sensor. The $M$ photons will divide into $N$ pairs of oppositely located spots, with the $n$th pair of spots corresponding to an obliquely propagating wave pair that carries the $Z_n$ projection of the incident wavefront  along the spherical-angle pair, $(\theta_n,\pm\phi_n)$, with $\theta_n=\sin^{-1}(q_n/k),\ \phi_n=\tan^{-1}(q_{ny}/q_{nx})$. The numbers of photons detected at the central pixels of the spots taken pairwise furnish estimates of the probabilities of the wavefront being in the corresponding modes. The remaining photons that are not detected provide an estimate of the wavefront being in the remaining states of a complete basis of which the subset, $\{Z_n,\ n=1,\ldots, N\},$ defines the observed states. 
The probability of detecting $m_1,\ldots, m_N$ photons in the $N$ projective channels is given by the multinomial (MN) distribution \cite{supp}, 
\be
\label{MN}
\Prob({\bar m},\{m_n\}|\{P_n\}) = M!{{\bar P}^{\bar m}\over \bar m!}\prod_{n=1}^N {(P_n)^{m_n}\over m_n!},
\ee
in which $\bar m=M-\sum_{n=1}^N m_n$ and $\bar P=1-\sum_{n=1}^N P_n$ are, respectively, the number and probability of undetected photons. Expressing the $P_n$ in terms of the separation coordinates, $l_x,l_y,l_z$, we performed their maximum-likelihood (ML) estimation by numerically minimizing $-\ln \Prob$  over those coordinates 
using Matlab's {\it fminunc} minimizer, which we started with an initial guess of $l_x=l_y=l_z=1/4$,  for a number of separations, 20,000 frames of noisy data, each with $M=10^6$ photons and generated using Matlab's {\it mnrnd} code.

\begin{figure}
\centerline{\includegraphics[width=1\columnwidth]{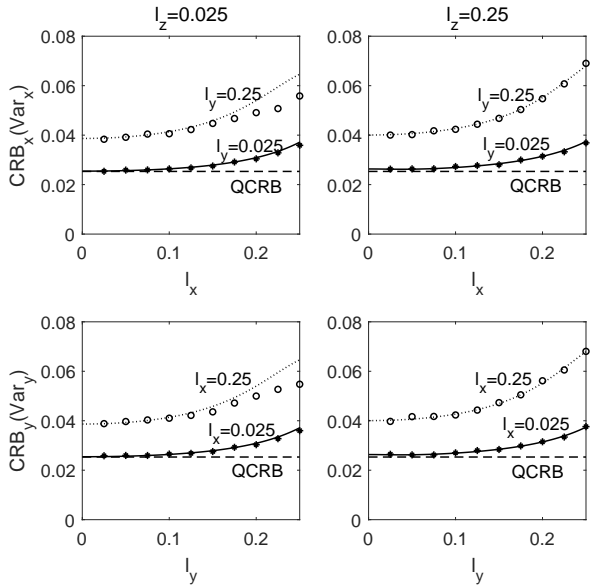}}
\vspace{-0.2cm}
\caption{Plots of CRBs w.r.t. $l_x$ for $l_y=0.025$ (lower curve) and $l_y=0.25$ (upper curve) and for $l_z= 0.025$ (left panels) and  $l_z=0.25$ (right panels).  The roles of $l_x$ and $l_y$ are interchanged in the bottom panels. Estimation variances obtained from simulation are shown by different marker symbols. }
\label{CRB1122projective}
\end{figure}

We plot in Fig.~\ref{CRB1122projective} the per-photon CRBs w.r.t. $l_x$ (top panels) and $l_y$ (bottom panels) for two different values of their axial separation, $l_z=0.025$ (left panels) and 0.25 (right panels). For each plot, we considered two different values, 0.025 and 0.25, of the other transverse coordinate, shown via the two different curves in each figure. Note that CRB w.r.t. each transverse-separation coordinate increases with increasing value of the other coordinate due to a cross-talk between the two transverse coordinates. Changing the longitudinal separation, however, has a less pronounced effect on those curves. As the pair separation increases, using only the first four Zernikes is insufficient to estimate $\blperp$, which accounts in part for the rising CRB curves.The discrete points identified by marker symbols are the results of the sample-based variance (per photon) of the ML estimate of the separation coordinates that we obtained in our numerical simulations. Note that the results of simulation are consistently lower than the corresponding CRB curves, which is most discernible in the left panels ($l_z=0.025$).  This is because the ML estimates of the separation coordinates are biased, particularly that for $l_z$, and standard CRBs do not provide the correct lower bounds without including bias-gradient based modifications \cite{VT68,Kay93}.

\begin{figure}
\centerline{\includegraphics[width=0.75\columnwidth]{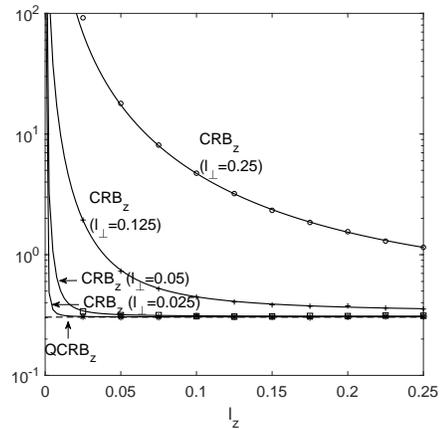}}
\caption{Plots of CRB w.r.t. $l_z$, for four different values of $l_\perp$, namely 0.025, 0.05, 0.125, and 0.25. Simulated estimation variances are shown by different marker symbols. }
\label{CRB33projective}
\end{figure}

In Fig. \ref{CRB33projective} we plot the per-photon CRBs w.r.t. $l_z$ for four different values of $l_\perp$. We observe divergent behavior as $l_z$ approaches zero, corresponding to the vanishing of $J_{zz}[Z]$ whenver $l_\perp\neq 0$ that we noted earlier. 
This behavior is quite in contrast with the rather muted dependence on $l_z$ which we observed in Fig. \ref{CRB1122projective} for the CRBs w.r.t. $\blperp$. 
The cross-talk between the uncertainties in simultaneously estimating the three pair-separation coordinates inherently present in the small set of Zernike projections increases the CRB for $l_z$ estimation as $l_\perp$ increases. 
The simulated values of the variance for estimating $l_z$, indicated by marker symbols, agree well with the theoretical CRB values, with evidence of any bias only for $l_z<<1$.

This Letter has treated the fundamental error in estimating the full 3D separation vector for a source pair by calculating the corresponding QFI and proposing specific projection bases for which QFI is attainable. Simulations using the Zernike basis confirm our theoretical assertions.   

\section*{Acknowledgments}
The work was partially supported by the US Air Force Office of Scientific Research under grant no. FA9550-15-1-0286. The authors are grateful to G. Adesso for pointing out his group's very recent work \cite{Napoli18} on the simultaneous estimation of the angular and axial separations as well as the coordinates of the centroid of an incoherent source pair located in a single meridional plane. 

\end{document}


\title{Supplemental Material}

\author{Zhixian Yu and Sudhakar Prasad}
\date{\today}
\maketitle








\section{Derivation of QFI}\label{sec:QFIderivation}

By taking the matrix element of the SLD of $\hat\rho$ defined by the relation, 
\be
\label{SLD1}
\partial_\mu \hat \rho = {1\over 2}(\hat L_\mu\hat\rho +\hat\rho \hat L_\mu),
\ee 
between the eigenstates $|e_i\rangle$ and $|e_j\rangle$ of $\hat\rho$, with eigenvalues $e_i,e_j$, respectively, and dividing both sides of the resulting expression by $(e_i+e_j)/2$, we obtain  
\be
\label{SLD2}
\langle e_i|\hat L_\mu|e_j\rangle={2\langle e_i|\partial_\mu\hat\rho|e_j\rangle \over (e_i+e_j)}.
\ee
Use of this expression in the trace, $H_{\mu\nu}={\rm Tr}(\hat\rho\hat L_\mu\hat L_\nu)$, when the latter is evaluated in the eigenbasis of $\hat\rho$, immediately yields expression (8) of the main text.

For the two-state $\hat \rho$ being discussed in the paper, the general expression for the QFI matrix element reduces to the form,
\ba
\label{Hmn1}
H_{\mu\nu}&=\sum_{i=\pm}{4\over e_i}\langle e_i|\partial_\mu \hat\rho\partial_\nu \hat\rho|e_i\rangle\nn
&+\sum_{i=\pm}\sum_{j=\pm}\left[{4e_i\over {(e_i+e_j)}^2}-{4\over e_i}\right]\langle e_i|\partial_\mu \hat\rho|e_j\rangle\langle e_j|\partial_\nu \hat\rho|e_i\rangle.
\end{align}
We may simplify the derivatives in Eq.~(\ref{Hmn1}) by noting the eigenvector identity, $\partial_\mu [(\hat\rho-e_i\hat I)|e_i\rangle]=0,$ {\it i.e.},
\be
\label{der1}
\partial_\mu\hat\rho|e_i\rangle = \partial_\mu e_i|e_i\rangle -(\hat\rho-e_i\hat I)\partial_\mu|e_i\rangle.
\ee
Taking the inner product of Eq.~(\ref{der1}) with $\langle e_j|$ and then using the eigenrelation, $\langle e_j|(\hat\rho-e_i) = (e_j-e_i) \langle e_j|$, and the orthonormality of the eigenstates, we obtain one of the needed matrix elements,
\be
\label{der2}
\langle e_j|\partial_\mu\hat\rho|e_i\rangle = \delta_{ij}\partial_\mu e_i +(e_i-e_j)\langle e_j|\partial_\mu|e_i\rangle.
\ee
Multiplying Eq.~(\ref{der1}) by its adjoint, with $\mu$ in the former replaced by $\nu$, we obtain the following expression for the first of the matrix elements in Eq.~(\ref{Hmn1}):
\be
\label{der3}
\langle e_i|\partial_\mu \hat\rho\partial_\nu \hat\rho|e_i\rangle= \partial_\mu e_i\partial_\nu e_i +\partial_\mu \langle 
e_i|(\hat\rho-e_i\hat I)^2\partial_\nu|e_i\rangle,
\ee
with the eigenrelation, $(\hat\rho -e_i\hat I)|e_i\rangle=0$, eliminating the other two terms in the product. A substitution of relations (\ref{der2}) and (\ref{der3}) into Eq.~(\ref{Hmn1}) simplifies it, particularly when the $i=j$ terms in the double sum in Eq.~(\ref{Hmn1}) are combined with its first sum and we note that in the remaining two, $i\neq j$ terms of the double sum, $e_i+e_j=e_++e_-=1$, and $e_+-e_-=\Delta$. The QFI matrix element may thus be expressed as
\be
\label{Hmn2}
H_{\mu\nu}=\sum_{i=\pm}{1\over e_i}\partial_\mu e_i\partial_\nu e_i+4\sum_{i=\pm}{1\over e_i}(\partial_\mu \langle 
e_i|)(\hat\rho-e_i\hat I)^2\partial_\nu|e_i\rangle\nn
+4\Delta^2\sum_{i\neq j}\left({1\over e_i}-e_i\right)\langle e_i|\partial_\mu |e_j\rangle\langle e_j|\partial_\nu |e_i\rangle.
\ee

The eigenvalues and eigenstates of $\hat \rho$ have the specific values,
\be
\label{eigen}
e_\pm = (1\pm \Delta)/ 2;\ \ |e_\pm\ra =[2(1\pm \Delta)]^{-1/2}\left(|K_+\ra\pm |K_-\ra\right),\ \ \ \Delta=\la K_-|K_+\ra,
\ee
from which we see that 
\be
\label{dstate}
\partial_\mu |e_\pm\rangle = \mp {\partial_\mu \Delta\over 2(1\pm\Delta)}|e_\pm\rangle + {1\over \sqrt{2(1\pm\Delta)}}\left(\partial_\mu |K_+\rangle\pm \partial_\mu|K_-\rangle\right).
\ee
Taking the inner product of expression (\ref{dstate}) with $|e_+\rangle$ and $|e_-\rangle$ successively yields the matrix elements,
\ba
\label{MatEl1}
\langle e_+|\partial_\mu|e_+\rangle =& -{\partial_\mu\Delta\over 2(1+\Delta)}+{1\over 2(1+\Delta)} 
 (\langle K_+|+\langle K_-|)(\partial_\mu |K_+\rangle +\partial_\mu |K_-\rangle);\nn 
\langle e_-|\partial_\mu|e_+\rangle = &{1\over 2\sqrt{1-\Delta^2}} (\langle K_+|-\langle K_-|)(\partial_\mu |K_+\rangle +\partial_\mu |K_-\rangle).
\end{align}
Since $\langle K_+|K_+\rangle$ is 1, its derivative vanishes,
\be
\label{Identity1}
(\partial_\mu \langle K_+|)|K_+\rangle + \langle K_+|\partial_\mu|K_+\rangle=0.
\ee
But since the wavefunctions, $\langle\bu|K_\pm\rangle$, are of form, $P(\bu) \exp[\pm i (\phi_0-\Psi)]$, as defined in the main text, with only the phases $\phi_0,\Psi$ depending on the separation vector, $\bl$, we may express the first term on the LHS of Eq.~(\ref{Identity1}) as $\langle K_-|\partial_\mu|K_-\rangle$, which yields the following identity:
\be
\label{Identity2}
\langle K_+|\partial_\mu|K_+\rangle =-\langle K_-|\partial_\mu |K_-\rangle.
\ee
A similar argument applied to the pair of identities, $\partial_\mu(\langle K_-|K_+\rangle)=\partial_\mu (\langle K_+|K_-\rangle)=\partial_\mu \Delta$, in which $\Delta$ was chosen to be real by a proper choice of the phase constant, $\phi_0$, yields a second useful pair of identities,
\be
\label{Identity3}
\langle K_-|\partial_\mu|K_+\rangle  =\langle K_+|\partial_\mu|K_-\rangle  ={1\over 2}\partial_\mu \Delta.
\ee
By applying results (\ref{Identity2}) and (\ref{Identity3}), we may simplify expressions (\ref{MatEl1}) greatly,
\be
\label{MatEl2}
\langle e_+|\partial_\mu|e_+\rangle =0;\ \ 
\langle e_-|\partial_\mu|e_+\rangle = {1\over \sqrt{1-\Delta^2}} \langle K_+|\partial_\mu |K_+\rangle.
\ee

In view of the eigen-relation,
$$\hat\rho =e_+|e_+\rangle\langle e_+|+e_-|e_-\rangle\langle e_-|, 
$$
and the first of the relations (\ref{MatEl2}), we may write
\be
\label{States2}
 (\hat\rho-e_+\hat I)\partial_\mu |e_+\rangle=e_-|e_-\rangle\langle e_-|\partial_\mu |e_+\rangle - e_+\partial_\mu|e_+\rangle.
\ee
The inner product of state (\ref{States2}) with its adjoint yields
\ba
\label{MatEl4}
(\partial_\mu 
\langle e_+|)(\hat\rho-e_+\hat I)^2\partial_\nu|e_+\rangle
&=(e_-^2-e_-e_+)\langle e_-|\partial_\mu |e_+\rangle^* \langle e_-|\partial_\nu |e_+\rangle\nn
&-e_-e_+(\partial_\mu\langle e_+|)|e_-\rangle\langle e_-|\partial_\nu |e_+\rangle
+e_+^2(\partial_\mu\langle e_+|)\partial_\nu|e_+\rangle\nn
&=-(e_-^2-2e_-e_+)\langle e_+|\partial_\mu |e_-\rangle \langle e_-|\partial_\nu |e_+\rangle+e_+^2(\partial_\mu\langle e_+|)\partial_\nu|e_+\rangle,
\end{align}
in which the relation, $(\partial_\mu\langle e_+|)|e_-\rangle=\partial_\mu(\langle e_+|e_-\rangle)-\langle e_+|\partial_\mu |e_-\rangle$, in conjunction with $\langle e_+|e_-\rangle=0$, was used to arrive at the last equality. 

By substituting formula (\ref{dstate}) for the derivatives of the eigenstates, we may calculate the last matrix element in Eq.~(\ref{MatEl4}) as
\ba
\label{MatEl5}
(\partial_\mu \langle e_+|)\partial_\nu|e_+\rangle
=&\left[-{\partial_\mu \Delta\over 2(1+\Delta)}\langle e_+|+
{1\over\sqrt{2(1+\Delta)}}\left(\partial_\mu\langle K_+|+\partial_\mu\langle K_-|\right)\right]\nn
&\times \left[-{\partial_\nu \Delta\over 2(1+\Delta)} |e_+\rangle+
{1\over\sqrt{2(1+\Delta)}}\left(\partial_\nu|K_+\rangle +\partial_\nu|K_-\rangle\right)\right]\nn
=&-{\partial_\mu\Delta\,\partial_\nu\Delta\over 4(1+\Delta)^2}
+{1\over 2(1+\Delta)}\left(\partial_\mu\langle K_+|+\partial_\mu\langle K_-|\right)
\left(\partial_\nu|K_+\rangle +\partial_\nu|K_-\rangle\right),
\end{align}
in which to arrive at the last equality we used expression (\ref{eigen}) for the eigenstate $|e_+\rangle$ and the first of the identities in Eq.~(\ref{MatEl2}) and its Hermitian adjoint to make the simplifications,
\be
\label{identity3}
 \langle e_+|(\partial_\nu |K_+\rangle+\partial_\nu|K_-\rangle)={1\over \sqrt{2(1+\Delta)}}
 \partial_\nu \Delta;\ \
(\partial_\mu \langle K_+|+\partial_\mu\langle K_-|)|e_+\rangle={1\over \sqrt{2(1+\Delta)}}
 \partial_\mu \Delta.
 \ee
 
Interchanging $e_+$ and $e_-$ everywhere in Eq.~(\ref{MatEl4}) yields the second matrix element we need,
 \be
\label{MatEl6}
(\partial_\mu 
\langle e_-|)(\hat\rho-e_-)^2\partial_\nu|e_-\rangle
=-(e_+^2-2e_-e_+)\langle e_-|\partial_\mu |e_+\rangle \langle e_+|\partial_\nu |e_-\rangle
+e_-^2(\partial_\mu\langle e_-|)\partial_\nu|e_-\rangle,
\ee
in which the last of the matrix elements is given by replacing $\Delta$ by $-\Delta$ and $|K_-\rangle$ by $-|K_-\rangle$ in Eq.~(\ref{MatEl5}),
\be
\label{MatEl7}
(\partial_\mu \langle e_-|)\partial_\nu|e_-\rangle
=-{\partial_\mu\Delta\,\partial_\nu\Delta\over 4(1-\Delta)^2}
+{1\over 2(1-\Delta)}\left(\partial_\mu\langle K_+|-\partial_\mu\langle K_-|\right)
\left(\partial_\nu|K_+\rangle -\partial_\nu|K_-\rangle\right).
\ee

 Since  $K_\pm(\bu)$ are mutually complex-conjugate phase exponentials over the aperture, it follows that $(\partial_\mu \langle K_+|)\partial_\nu|K_-\rangle =(\partial_\mu\langle K_-|)\partial_\nu |K_+\rangle^*$ and $(\partial_\mu  \langle K_+|)\partial_\nu|K_+\rangle =(\partial_\mu\langle K_-|)\partial_\nu |K_-\rangle^*$,  the latter being already real, the last part of expression (\ref{MatEl5}) reduces further. Substituting the so-reduced from of this expression into relation (\ref{MatEl4}) and the resulting expression into form (\ref{Hmn2}) for the QFI matrix element and noting from relation (\ref{eigen}) that $\partial_\mu e_i\partial_\mu e_i=(1/4)\partial_\mu \Delta\partial_\nu\Delta$, $i=\pm$, yield an exact cancellation of all $\partial_\mu\Delta\partial_\nu\Delta$ terms and yield the following simplified expression for the QFI matrix element:
\begin{align}
\label{Hmn3}
H_{\mu\nu}
&=-{4\over e_+}\left\{ (e_-^2-2e_-e_+)\langle e_+|\partial_\mu |e_-\rangle \langle e_-|\partial_\nu |e_+\rangle\right\}
-{4\over e_-}\left\{ (e_+^2-2e_-e_+)\langle e_-|\partial_\mu |e_+\rangle \langle e_+|\partial_\nu |e_-\rangle\right\}\nn
&+4(\partial_\mu  \langle K_+|)\partial_\nu|K_+\rangle] +4\Delta^2\sum_{i\neq j}\left({1\over e_i}-e_i\right)\langle e_i|\partial_\mu |e_j\rangle\langle e_j|\partial_\nu |e_i\rangle,\nn
\end{align} 
in which we performed some additional simplifications to achieve the last equality. The first two terms and the last term on the RHS of Eq.~(\ref{Hmn3}) may be combined and simplified by noting that $e_-^2-2e_+e_-=(e_+-e_-)^2-e_+^2=\Delta^2-e_+^2$ and analogously $e_+^2-2e_+e_-=\Delta^2-e_-^2$ to derive the more compact result,
\begin{align}
\label{Hmn4}
H_{\mu\nu}
&=4\left[(\partial_\mu  \langle K_+|)\partial_\nu|K_+\rangle+(1-\Delta^2)\langle e_+|\partial_\mu |e_-\rangle\langle e_-|\partial_\nu |e_+\rangle\right]\nn
&=4\left[(\partial_\mu  \langle K_+|)\partial_\nu|K_+\rangle+\langle K_+|\partial_\mu |K_+\rangle \langle K_+|\partial_\nu|K_+\rangle\right],
\end{align} 
in which the second identity in Eq.~(\ref{MatEl2}) was used to reach the last equality.  By noting the relation,
\be
\label{Phase}
\partial_\mu \langle \bu|K_+\rangle = i[\partial_\mu \phi_0-\partial_\mu \Psi(\bu)]\langle \bu|K_+\rangle, 
\ee
which follows from the form of the wavefunction, $\la \bu|K_+\ra\sim P(\bu) \exp[i\phi_0-i\Psi(\bu,\bl)]$, we may reduce expression (\ref{Hmn4}) to its simplest final form,
\be
\label{Hmn5}
H_{\mu\nu}
=4\left[\langle \partial_\mu\Psi(\bu)\partial_\nu\Psi(\bu)\rangle-\langle \partial_\mu\Psi(\bu)\rangle\langle\partial_\nu\Psi(\bu)\rangle\right],
\ee 
where the angular brackets here denote weighted averages, with the weight function being $|P(\bu)|^2$, of quantities they enclose. 

\section{Some Properties of Sine and Cosine States}
\subsection{Orthonormaility and Completeness}
These states were defined as
\be
\label{basis1}
\begin{array}{ll}
\CC_{mn}(\bu)=\sqrt{c_m c_n\over \pi}\cos (2\pi m u^2)\cos n\phi,  & m,n=0,1,\ldots; \\
\CS_{mn}(\bu)=\sqrt{c_m c_n\over \pi}\cos (2\pi m u^2)\sin n\phi,  & m=0,1,\ldots,
\ n=1,2,\ldots; \\
\SC_{mn}(\bu)=\sqrt{c_m c_n\over \pi}\sin (2\pi m u^2)\cos n\phi,  & m=1,2,\ldots,
\ n=0,1,\ldots; \\
\SS_{mn}(\bu)=\!\sqrt{c_m c_n\over \pi}\sin (2\pi m u^2)\sin n\phi,  & m,n=1,2,\ldots;
\end{array}
\ee
in which the normalization constant, $c_n$, has the value, $c_n=2-\delta_{n0}$.
Denoting the most general of these basis functions simply as $A_{mn}(\bu)=\langle \bu|A_{mn}\rangle$, we can, by standard trigonometric integrations, easily prove their orthonormality over the unit disk,
\be
\label{ortho}
\int_0^1 du\, u\int_0^{2\pi}d\phi A_{mn}^*(\bu)\, A_{m^\prime n^\prime}(\bu) = \delta_{mm^\prime}\delta_{nn^\prime}.
\ee
Their completeness,
\be
\label{complete}
\sum_{m,n=0}^\infty\sum_{\substack{A=\CC,\\ \CS, \SC,\SS}} A_{mn}^*(\bu)A_{mn}(\bw)=\delta^{(2)}(\bu-\bw),
\ee
follows from the Poisson summation formulas involving sums over non-negative integer values of $m,n$, 
\be
\label{poisson}
\sum_m c_m\cos 2\pi m(v-w)\!=\delta(v-w);\ \ 
\sum_n c_n\cos n(\phi-\psi)=2\pi\delta(\phi-\psi);
\ee
valid over the unit disk, $0\leq v,w\leq 1;\  0\leq \phi,\psi<2\pi$. 

\subsection{The Overlap Integrals $\la A_{mn}|K_\pm\ra$, $A=\CC,\CS,\SC,\SS$}
For the choice, $\phi_0=0$, of the phase constant of the photon wavefunctions, the overlap integrals, $\la A_{mn}|K_\pm\ra$,  for a transversely separated source pair, $\blperp\neq 0, \ l_z=0$, and a clear aperture, are given by
\be
\label{overlap}
\la A_{mn}|K_\pm\ra={1\over \sqrt{\pi}}\int_0^1 du\, u\oint d\phi \exp(\mp i2\pi \bu\cdot\blperp)\, A_{mn}(\bu).
\ee
Since $\bu\cdot\blperp=u\, l_\perp\cos(\phi-\phi_l)$, in which $\phi_l$ is the polar angle of $\blperp$, the following integral identities are easily proved using the Bessel-function generating function formula:
\ba
\label{identity2}
\oint d\phi \exp[\mp iz\cos(\phi-\phi_l)]\cos n\phi=&(\mp i)^n 2\pi\cos n\phi_l\, J_n(z),\nn  
\oint d\phi \exp[\mp iz\cos(\phi-\phi_l)]\sin n\phi=&(\mp i)^n 2\pi\sin n\phi_l\, J_n(z),
\end{align}
in which $J_n(z)$ denotes the ordinary Bessel function of order $n$. Use of these identities allows us to perform the $\phi$ integral in Eq.~(\ref{overlap}), 
We thus reduce all of the probabilities to simple integrals over a single convenient radial variable, $v=u^2$,
\ba
\label{prob4}
\la \CC_{mn}|K_\pm\ra=&(\mp i)^n \sqrt{c_mc_n}\cos n\phi_l\int_0^1 dv\cos (2\pi m v) J_n(2\pi l_\perp\sqrt{ v}); \nn
\la \CS_{mn}|K_\pm\ra=&(\mp i)^n \sqrt{c_mc_n}\sin n\phi_l\int_0^1 dv\cos (2\pi m v) J_n(2\pi l_\perp\sqrt{ v}); \nn
\la \SC_{mn}|K_\pm\ra=&(\mp i)^n \sqrt{c_mc_n}\cos n\phi_l\int_0^1 dv\sin (2\pi m v) J_n(2\pi l_\perp\sqrt{ v}); \nn
\la \SS_{mn}|K_\pm\ra=&(\mp i)^n \sqrt{c_mc_n}\sin n\phi_l\int_0^1 dv\sin (2\pi m v) J_n(2\pi l_\perp\sqrt{ v});
\end{align}
which are all real integrals whose phases are either 0 or $\pi$ ({\rm mod} $2\pi$), which are constants independent of $\blperp$, and whose magnitudes satisfy the relation, $|\la A_{mn}|K_+\ra|=\la A_{mn}|K_-\ra|$, $A=\CC,\CS,\SC,\SS$. Because of these two properties, this complete basis achieves QFI for a transversely separated source pair.

\section{Probabilities of Zernike Projections}

The probability of the photon wave function being in the Zernike mode, $Z_n$, may be expressed as
\be
\label{Zprob1}
P_n = \left|\int d^2u \,P(\bu)\, Z_n(\bu)\,\cos\Psi\right|^2
       +\left|\int d^2u \,P(\bu)\, Z_n(\bu)\,\sin\Psi\right|^2,
\ee
with the probability of finding it in the remaining, unmeasured modes being
\be
\label{Zprob2}
\bar P=1-P_1-P_2-P_3-P_4.
\ee
For a clear circular aperture, for which $P(\bu)$ is simply $1/\sqrt{\pi}$ times the indicator function of the unit-radius aperture, and for small separation coordinates, $l_\perp,l_z<<1$, 
we retain only the first two orders in the Taylor expansions of the $\sin\Psi$ and $\cos\Psi$ in Eq.~(\ref{Zprob1}). Orthonormality of the Zernikes implies $\int d^2 u \, P(u)\, Z_n(\bu)=\delta_{n1}$, from which it follows that up to the lowest two orders in $\Psi$ and thus in $\bl$, $P_n$ has the form,
\be
\label{prob9}
P_n=\left\{
\begin{array}{ll}
1-(\la\Psi^2\ra-\la\Psi\ra^2) & n=1\\
\pi \left[\la Z_n\Psi\ra^2+{1\over 4}\la Z_n\Psi^2\ra^2-{1\over 3} \la Z_n\Psi\ra\la Z_n\Psi^3\ra\right] & n\geq 2,
\end{array}
\right.
\ee 
in which angular brackets denote averages over the clear aperture. Using expressions for the wavefront phase,
\be
\label{Phase}
\Psi(\bu;\bl) = 2\pi \blperp\cdot \bu +\pi l_z u^2,
\ee
and the Zernike modes, 
\ba
\label{Zernikes}
Z_1=&{1\over \sqrt{\pi}},\  Z_2={2 \over\sqrt{\pi}}u\cos\phi,\ Z_3= {2 \over\sqrt{\pi}}u\sin\phi,\nn
Z_4=&\sqrt{3\over\pi} (2u^2-1),
\end{align}
we may easily evaluate these averages to obtain the probabilities to two lowest significant orders in the separation vector, $\bl$, 
\be
\label{prob10}
P_n=\left\{
\begin{array}{ll}
1-\pi^2(l_\perp^2+l_z^2/12)+O(l_\perp^4,l_z^4) & n=1\\
\pi^2 l_x^2[1+O(l_\perp^2,l_z^2)] & n=2\\
\pi^2 l_y^2 [1+O(l_\perp^2,l_z^2)]& n=3\\
\pi^2 l_z^2/12 +O(l_z^4,l_z^2l_\perp^2,l_\perp^4)& n=4
\end{array}
\right.
\ee

\section{Likelihood Function for Photon Division into $N$ Channels}
Consider division of $M$ photons into $N$ channels, with $P_k$ being the probability of a photon going into the $k$th channel.  If $n_k$ is the number of photons transmitted into the $k$th channel in a statistical realization of this process, then the probability of this process is given by the multinomial (MN) distribution,
\be
\label{prob1}
\Prob(\{n_k\}|M)=M!\prod_{k=1}^N {P_k^{n_k}\over n_k!}, 
\ee
in which all photon numbers, $n_1,\ldots, n_N$ and $\bar n$ are non-negative and thus each bounded above by $M$. Let $\eta$ be the quantum efficiency (QE) of detection of the transmitted photons in each channel, then the probabilty of detection of $m_k$ photons in the $k$th channel , $k=1,\ldots, N$, conditioned on the knowledge that $\{n_k,\ k=1,\ldots, N\}$ photons were transmitted into the various channels, is given by a product of binomial distributions,
\be
\label{prob2}
\Prob(\{m_k\}|\{n_k\},M)=\prod_{k=1}^N {n_k!\over m_k!(n_k-m_k)!}\eta^{m_k}(1-\eta)^{n_k-m_k}. 
\ee
The probability of jointly detecting $m_k$ photons in the $k$th channel, with $k=1,2,\ldots, N$, is then given by the composition rule,
\ba
\label{Prob3}
\Prob(\{m_k\}|M)=&\sum_{\substack{\{n_k\in(m_k,\ldots,M)\}\\
                                                      \sum_k n_k= M}}\Prob(\{m_k\}|\{n_k\},M)\Prob(\{n_k\}|M) \nn
                                       =&M!\sum_{\substack{\{n_k\in(m_k,\ldots,M)\}\\
                                                      \sum_k n_k= M}}\prod_{k=1}^N \left[{P_k^{n_k}(1-\eta)^{n_k}\over (n_k-m_k)!}                 
                                                      {\eta^{m_k}(1-\eta)^{-m_k}\over m_k!}\right]\nn
                                       =&M!\left[\prod_{k=1}^N {(\eta P_k)^{m_k}\over m_k!}\right]\prod_{\substack{k=1\\ \sum_k \delta_k=M-\sum_k m_k}}^N\left[ \sum_{\delta_k=0}^{M-m_k}{[P_k(1-\eta)]^{\delta_k}\over \delta_k!}\right],
\end{align}
in which the transformation, $\delta_k=n_k-m_k$, was used to replace the sum over $n_k$ to that over $\delta_k$. The latter product of the sums, with the restriction that the sum of the values of the indices $\delta_k$ be constrained to be a fixed number, can be performed by using the following identity involving the product of exponentials:
\ba
\label{exp_id}
\exp\left[(1-\eta)\sum_{k=1}^N P_k\right]=&\prod_{k=1}^N \exp[(1-\eta) P_k)]\nn
                                        =&\prod_{k=1}^N \sum_{\delta_k=0}^\infty {[P_k(1-\eta)]^{\delta_k}\over \delta_k!},
\end{align}
and noting that its left-hand side may be expanded in powers of $(1-\eta)$. Comparing the $(1-\eta)^\delta$ term on both sides of the resulting identity then yields the needed relation,
\ba
\label{ident6}
\prod_{\substack{k=1\\ \sum_k \delta_k=\delta}}^N\left[ \sum_{\delta_k=0}^{M-m_k}{[P_k(1-\eta)]^{\delta_k}\over \delta_k!}\right]=&{\left[\sum_{k=1}^N P_k(1-\eta)^\delta\right]\over \delta!}\nn
               =&{(1-\eta)^\delta\over \delta!},
\end{align}
since the probabilities $P_k$ sum to 1 over all $N$ channels. When relation (\ref{ident6}), with $\delta$ replaced by $M-\sum_k m_k$, is substituted into expression (\ref{Prob3}), we can simplify the latter to the form,
\be
\label{Prob4}
\Prob(\{m_k\}|M) =M!{(1-\eta)^{M-\sum_k m_k}\over (M-\sum_k m_k)!}\left[\prod_{k=1}^N {(\eta P_k)^{m_k}\over m_k!}\right],
\ee
which has a very compelling interpretation that non-unit QE provides yet another channel, the $(N+1)$th channel, which ``captures'' the undetected counts, while the other channels capture photons at the compounded probabilities, $\eta P_k$, per photon for the $k$th channel, with $k=1,\ldots,M$.

Note that for a given set of detected counts, $\{m_1,\ldots,m_N\}$, the probability (\ref{Prob4}) reduces to a product of a fixed $\eta$ dependent factor and another that depends on the per-photon channel probabilities $P_k$, $k=1,\ldots,N$. This implies that the maximum-likelihood estimation of the latter probabilities from the likelihood function (\ref{Prob4}) is independent of $\eta$. For this reason, there is no loss of generality in choosing $\eta=1$. 

We must also interpret the $N$ modes in expression (\ref{Prob4}) as including the 4 Zernikes modes into which the wavefront is projected as well as the remaining modes into which the wavefront is not projected, with the latter to be regarded as a single undetected mode, which we denote by an overhead bar. In other words, for $\eta=1$, one must modify that expression to the form,  
\be
\label{Prob5}
\Prob(\{m_k\}|M) =M!{{\bar P}^{\bar m}\over {\bar m}!}\left[\prod_{k=1}^N {(\eta P_k)^{m_k}\over m_k!}\right],
\ee
with $\bar P = 1-\sum_{k=1}^N P_k$ and $\bar m=M-\sum_{k=1}^N m_k$.